\documentclass[12pt]{iopart}

\usepackage{bm}
\usepackage{amssymb}
\usepackage{amstext}
\usepackage{amsmath}
\usepackage{mathrsfs}
\usepackage{graphicx}
\usepackage{color}
\usepackage{amsthm}
\usepackage{floatrow}
\usepackage{subfig}
\usepackage{caption}
\usepackage{float}
\usepackage{array}

\usepackage{listings}
\usepackage{booktabs}
\usepackage{tikz}
\usepackage{hyperref}

\begin{document}

\title{Self-organized critical dynamics of RNA virus evolution}

\author{Xiaofei Ge$^{1}$}
\address{School of Pharmaceutical Sciences, Tsinghua University, Beijing, 100084, China.}
\ead{gxf20@mails.tsinghua.edu.cn$^{1}$}

\author{Kaichao You$^{2}$}
\address{School of Software, Tsinghua University, Beijing, 100084, China.}
\ead{ykc20@mails.tsinghua.edu.cn$^{2}$}

\author{Zeren Tan$^{3}$}
\address{Institute for Interdisciplinary Information Science, Tsinghua University, Beijing, 100084, China.}
\ead{tanzr20@mails.tsinghua.edu.cn$^{3}$}

\author{Hedong Hou$^{4}$}
\address{UFR de Math\'{e}matiques, Universit\'{e} de Paris, Paris, 75013, France.}
\ead{hedong.hou@etu.u-paris.fr$^{4}$}

\author{Yang Tian$^{5}$ and Pei Sun$^{6}$}
\address{Department of Psychology \& Tsinghua Laboratory of Brain and Intelligence, Tsinghua University, Beijing, 100084, China.}
\ead{tiany20@mails.tsinghua.edu.cn$^{5}$; peisun@tsinghua.edu.cn$^{6}$}

\vspace{10pt}
\begin{indented}
\item[]February 2021
\end{indented}

\begin{abstract}
RNA virus (e.g., SARS-CoV-2) evolves in a complex manner. Studying RNA virus evolution is vital for understanding molecular evolution and medicine development. Scientists lack, however, general frameworks to characterize the dynamics of RNA virus evolution directly from empirical data and identify potential physical laws. To fill this gap, we present a theory to characterize the RNA virus evolution as a physical system with absorbing states and avalanche behaviors. This approach maps accessible biological data (e.g., phylogenetic tree and infection) to a general stochastic process of RNA virus infection and evolution, enabling researchers to verify potential self-organized criticality underlying RNA virus evolution. We apply our framework to SARS-CoV-2, the virus accounting for the global epidemic of COVID-19. We find that SARS-CoV-2 exhibits scale-invariant avalanches as mean-field theory predictions. The observed scaling relation, universal collapse, and slowly decaying auto-correlation suggest a self-organized critical dynamics of SARS-CoV-2 evolution. Interestingly, the lineages that emerge from critical evolution processes coincidentally match with threatening lineages of SARS-CoV-2 (e.g., the Delta virus). We anticipate our approach to be a general formalism to portray RNA virus evolution and help identify potential virus lineages to be concerned.
\end{abstract}

%
%
%
%
%

\section{Introduction}
With high mutation and replication capacities, RNA viruses exhibit intricate evolution \cite{domingo1996basic,dolan2018mechanisms,duffy2018rna}. Studying the complex evolution dynamics of RNA virus remains as an attractive challenge in biology, as RNA virus evolution is a frequent cause of new pathogens and may be a pivotal route in medicines \cite{ojosnegros2010models,irwin2016antiviral,jakobsson2020combining}. Meanwhile, RNA virus interests physicians for its potential to serve as a representative model of long-term molecular evolution \cite{sneppen2005physics,tsimring1996rna}. 

Despite its apparent significance, the physics underlying RNA virus evolution remains elusive. While important progress has been accomplished in characterizing the dynamics of mutant virus population size (e.g., see Refs. \cite{fontana1989physical,tsimring1996rna,sasaki1994evolution,haraguchi1997evolutionary,sasaki2000antigenic,nowak2000virus,kauffman1993origins,korobeinikov2004global,regoes1998virus,adiwijaya2010multi,garcia2018clonal,yan2019phylodynamic,held2019survival,rasmussen2019coupling,marchi2021antigenic,korobeinikov2016multi}), there remains another non-negligible aspect of evolution for exploration. From the biological perspective, the dynamics accountable for phylogenetic tree formation is informative for the global evolution properties \cite{lassig2014adaptive,neher2014predicting,desai2007beneficial,rouzine2003solitary}. Compared with the evolution previously considered in the context of population dynamics \cite{fontana1989physical,tsimring1996rna,sasaki1994evolution,haraguchi1997evolutionary,sasaki2000antigenic,nowak2000virus,kauffman1993origins,korobeinikov2004global,regoes1998virus,adiwijaya2010multi,garcia2018clonal,yan2019phylodynamic,held2019survival,rasmussen2019coupling,marchi2021antigenic,korobeinikov2016multi}, this unexplored aspect may convey key information about the macroscopic laws of RNA virus evolution \cite{lassig2014adaptive,neher2014predicting,desai2007beneficial,rouzine2003solitary} but yet remains unclear. Moreover, existing models either non-trivially handle multiple time scales with extreme differences in magnitude orders or are simplified by phenomenological parameters \cite{korobeinikov2016multi}. Although phenomenological models are more solvable on empirical data than non-trivial analytic ones, these models are powerful in imitating phenomena but limited in explaining mechanisms. Interpretations of phenomenological parameters with experiments should be prudently considered \cite{korobeinikov2016multi}. In sum, much unknown remains about the general frameworks that are applicable to identify potential physical laws of RNA virus evolution from empirical data.

The objectives of the current research are twofold. We aim at exploring the global properties of RNA virus evolution by developing a physics theory of the dynamics underlying phylogenetic tree formation. The theory is demanded to map fundamental biological factors to general characterizations of system dynamics so as to guarantee its applicability on real data and universality to arbitrary RNA viruses. Applying this theory, we present a systematic analysis of SARS-CoV-2, the virus accounting for COVID-19 \cite{velavan2020covid}. The global infection of SARS-CoV-2 provides us a unique opportunity with rich data (e.g., a sufficiently large phylogenetic tree) to validate our theory. Moreover, confirming the macroscopic characteristics of SARS-CoV-2 evolution may benefit to developing coping strategies of COVID-19.

\section{Backgrounds of RNA virus}
To make our analysis comprehensible, we begin with a fundamental perspective of RNA virus. In RNA virus populations, the replication implies the vanishing of the original RNA virus and the birth of new RNA viruses that can also replicate themselves \cite{v2021coronavirus}. The proliferation of RNA virus is bounded by time-dependent host cell resource $\lambda\left(t\right)$ \cite{domingo1996basic,tsimring1996rna,bar2020science}. Specifically, the maximum cumulative mutation rate, $\sup_\omega\widehat{\zeta}\left(\omega\right)$, of an RNA virus population of initial size $\omega\in\mathbb{N}^{+}$ during a duration of $\left[0,\kappa\right]$  is limited by $\lambda\left(t\right)$ in the following way
\begin{align}
    \sup_{\lambda}\sup_\omega\widehat{\zeta}\left(\omega\right)=\omega\sigma\tau\Big(\phi^{\lfloor \kappa/\tau\rfloor-1}-\frac{1}{\phi}\Big),\label{EQ1}
\end{align}
where $\sigma$ is the mutation rate of single RNA virus, parameter $\tau$ denotes entrance and eclipse periods (assume $\tau$ is divisible by $\kappa$), and $\phi\simeq 10^{3}$ denotes burst size (the number of viruses produced by a host cell) \cite{domingo1996basic,tsimring1996rna,bar2020science}.
Here $\sup_\omega\widehat{\zeta}\left(\omega\right)$ can reach the maximum cumulative mutation rate of an RNA virus population under ideal conditions when host cell resources are ample $\lambda\left(t\right) \geq \phi^{\lfloor t/\tau\rfloor}-\phi^{\lfloor t/\tau\rfloor-1}$. Derivations of Eq. (\ref{EQ1}) are presented in \ref{AP1}.

In summary, host cell resource $\lambda\left(t\right)$ intrinsically determines the cumulative daily mutation rate of a RNA virus population. Instead of proposing detailed definitions of $\lambda\left(t\right)$ as previous studies (e.g., define life circles of cells) \cite{nowak2000virus,korobeinikov2004global,adiwijaya2010multi,yan2019phylodynamic,held2019survival,marchi2021antigenic,korobeinikov2016multi}, we leave the characterization of $\lambda\left(t\right)$ as a task of epidemic and observe the evolution dynamics of RNA virus controlled by $\lambda\left(t\right)$ directly from  empirical data.

\section{Theory of RNA virus evolution}
Certainly, one can not expect to count $\lambda\left(t\right)$ in a real epidemic. This parameter can be reasonably replaced by the number of human hosts. Assuming that a lineage $l_{i}$ of RNA virus emerges at moment $t^{\prime}$ and one of its progeny lineage $l_{j}$ is born at moment $t^{\prime\prime}$, we can measure the duration length of evolution as $T=t^{\prime\prime}-t^{\prime}$. Meanwhile, we define $S\left(T\right)$ as the number of cumulative confirmed patients with lineage $l_{i}$ during $\left[t^{\prime},t^{\prime\prime}\right]$ and denote the time-dependent number of confirmed patients with lineage $l_{i}$ by $S\left(t\mid T\right)$ (here $t\in\left[t^{\prime},t^{\prime\prime}\right]$). The terminology ``lineage" can be generally understood as a kind of definition of virus sub-type (e.g., see examples in Ref. \cite{rules}). Viruses of lineage $l_{i}$ need to accumulate sufficient mutations to make lineage $l_{j}$ emerge.

We can relate these concepts with a branching process concerning the number of copies of $l_{i}$ during $\left[t^{\prime},t^{\prime\prime}\right]$, that is, a process where a patient with lineage $l_{i}$ infects multiple people before the virus mutates to lineage $l_{j}$ during replication. The realization of this branching process, from the initialization $t^{\prime}$ to the termination $t^{\prime\prime}$ after which a new lineage $l_{j}$ emerges, is referred to as an avalanche. The size and life time of avalanche are exactly $S\left(T\right)$ and $T$. This branching process is slightly different from the standard one, where the termination is defined as a moment after which the population size (e.g., the number of patients with lineage $l_{i}$) remains $0$ indefinitely \cite{harris1963theory,garcia2018field}. 

We can further relate the branching process of evolution with directed percolation, a universality class exhibiting a phase transition separating an absorbing state from an active state \cite{harris1963theory,henkel2008non,hinrichsen2006non}. In directed percolation (e.g., the contact process \cite{harris1974contact}), sites (e.g., people) can be either active (infected) or inactive (healthy). With different balance between infection and recovery, the infection process may propagate over the system or vanish gradually. If infection vanishes, the system is trapped in a completely inactive state, the so-called absorbing state. Directed percolation is a non-equilibrium process since detailed balance breaks in the absorbing state (it can be reached but not be escaped) \cite{hinrichsen2000non}. 

We are interested in potential self-organized criticality (SOC) due to its pervasiveness in biological systems (e.g., neural avalanches in the brain \cite{fontenele2019criticality,pausch2020time}). The self-organization to criticality distinguishes SOC (e.g., Refs. \cite{bak1987self,manna1991two}) from ordinary critical phenomena \cite{hinrichsen2000non,lubeck2004universal}. At criticality, the system jumps between absorbing configurations by avalanches \cite{hinrichsen2000non,lubeck2004universal}. From the biological perspective, this property implies an unpredictable and unpreventable evolution of RNA virus. Numerous mean field theories of directed percolation have been proposed to determine avalanche exponents to study the scale-invariance underlying $\mathcal{P}_T$ and $\mathcal{P}_S$, the probability distributions of $T$ and $S$ (e.g., see Ref. \cite{zapperi1995self}). Note that $S$ is an abbreviation of $S\left(T\right)$. In \ref{AP2}, we re-calculate avalanche exponents following a similar idea of Refs. \cite{garcia1993branching,harris1963theory,otter1949multiplicative}. Our results show that
\begin{align}
   \mathcal{P}_T\left(t\right)\propto t^{-2},\label{EQ2}\\
   \mathcal{P}_S\left(s\right)\propto s^{-3/2},\label{EQ3}
\end{align}
consistent with other mean field theory predictions of self-organized criticality \cite{zapperi1995self,garcia1993branching}.

To verify whether the system is at criticality, we can consider a scaling relation \cite{sethna2001crackling,baldassarri2003average,hinrichsen2000non,lubeck2004universal} in Eq. (\ref{EQ4}), where $\mathcal{P}_T\left(t\right)\propto t^{-\alpha}$, $\mathcal{P}_S\left(s\right)\propto s^{-\beta}$, and $\langle S\rangle\left(T\right)\propto T^{\gamma}$.
\begin{align}
    \gamma=\frac{\alpha-1}{\beta-1},\label{EQ4}
\end{align}
Inserting $\alpha=2$ and $\beta=\frac{3}{2}$ into Eq. (\ref{EQ4}), we derive $\gamma=2$. Apart from that, more precise verification of criticality can be implemented according to the collapse shape \cite{baldassarri2003average,pausch2020time} and the slow and exponential decay of auto-correlation \cite{pausch2020time}. Please see \ref{AP3} for details of these approaches. 

With host cell resource $\lambda\left(t\right)$ as a clue, we have derived a directed percolation theory to characterize RNA virus evolution as a system with absorbing states and avalanche behaviors. A series of approaches are presented to verify the existence of self-organized criticality. 

\section{Evolution of SARS-CoV-2}
Applying this theory, we aim at presenting a systematic analysis of SARS-CoV-2, the virus that causes COVID-19 \cite{velavan2020covid}. We use an open-source database \cite{database} to acquire the necessary data. The criterion of lineage definition can be seen in \cite{rules}. Our data includes all lineages that emerge before 27 December 2021.

In the SARS-CoV-2 database \cite{database}, the probability for a patient to be infected with each lineage of virus is estimated with confidence intervals. By multiplying the midpoint of confidence interval with the total number of worldwide patients recorded by the World Health Organization (WHO) \cite{whodatabase}, we can estimate the accumulative and daily numbers of confirmed patients with each lineage of virus, corresponding to $S\left(T\right)$ and $S\left(t\mid T\right)$ of each lineage (see \textbf{Fig. 1a}). Similarly, we can estimate the upper and lower bounds of $S\left(T\right)$ and $S\left(t\mid T\right)$ based on the upper and lower bounds of confidence interval. 

\begin{figure}[b]%
\centering
\includegraphics[width=1\textwidth]{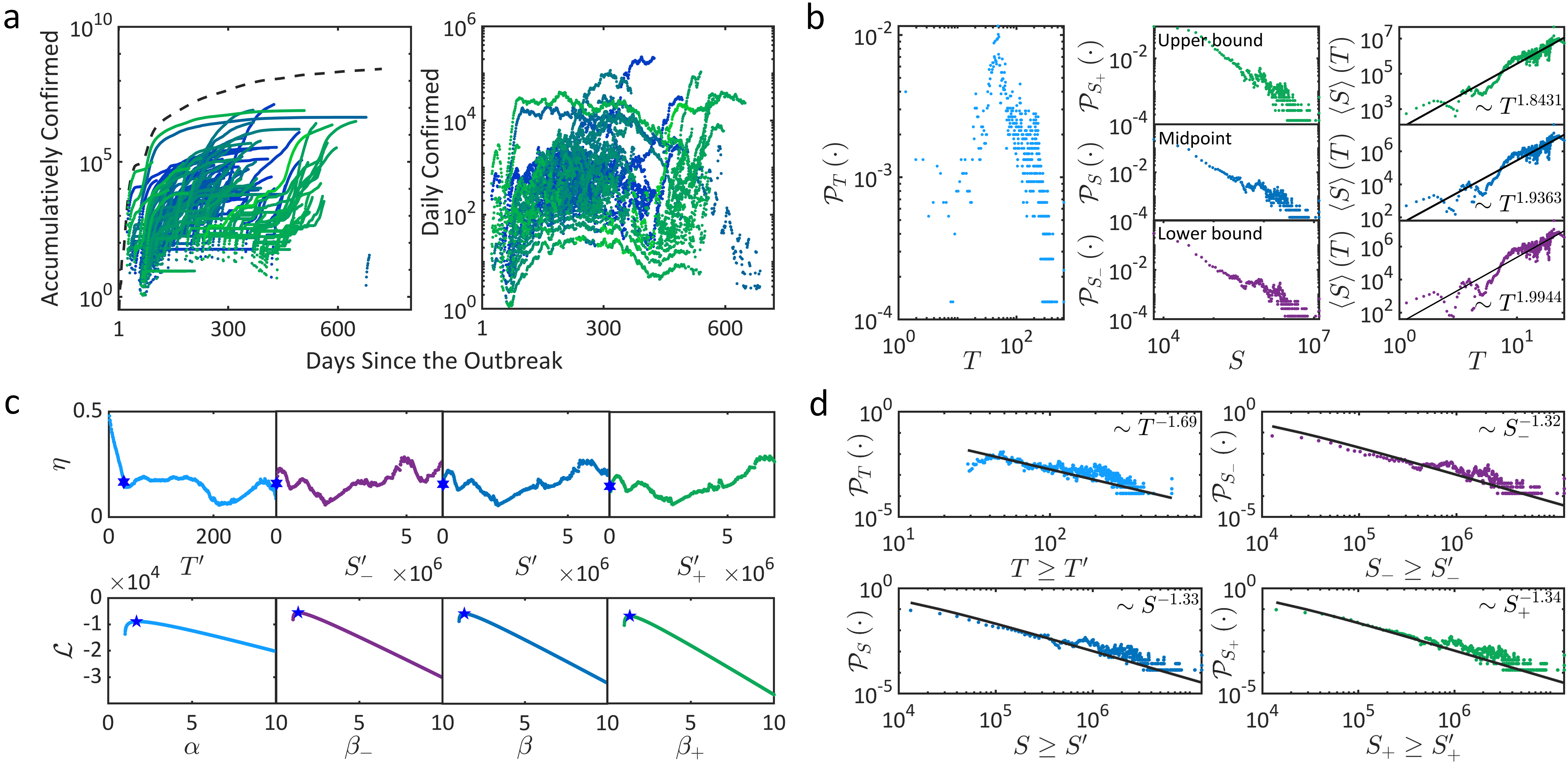}
\caption{Avalanches of SARS-CoV-2 evolution. \textbf{a} Numbers of accumulatively and daily confirmed patients with viruses of different lineages (scatters with different colors) and the total accumulative number of patients (black dashed line). \textbf{b} Data distributions of $\mathcal{P}_{T}\left(\cdot\right)$ vs. $T$, $\mathcal{P}_{S}\left(\cdot\right)$ vs. $S$, and $\langle S\rangle\left(T\right)$ vs. $T$. \textbf{c} The KS statistic $\eta$ for estimating distribution cutoffs and the likelihood $\mathcal{L}$ for estimating avalanche exponents. Blue stars denote estimated values. \textbf{d} The power-law distributions of $T$ and $S$ (midpoint and bounds of confidence interval) above cutoffs.}\label{fig1}
\end{figure}

In \textbf{Fig. 1b}, we show the probability distributions of life time $T$ and avalanche size $S$ (for midpoint and bounds). In real cases, the power-law may only hold on specific tails $\{\mathcal{P}\left(X\right)\vert X\geq X^{\prime}\}$ of the empirical distribution of variable $X$, where $X^{\prime}$ denotes a certain distribution cutoff \cite{clauset2009power}. We apply the approach in Ref. \cite{virkar2014power} to estimate cutoffs $T^{\prime}$, $S^{\prime}$, $S_{-}^{\prime}$ (lower bound), and $S_{+}^{\prime}$ (upper bound) as the points where the Kolmogorov-Smirnov (KS) statistic $\eta$ reduces to a sufficiently small threshold at the first time (\textbf{Fig. 1c}). Details of cutoff estimation can be seen in \ref{AP4}. Then, a maximum likelihood estimation of power-law exponent \cite{virkar2014power} is implemented on the samples above cutoffs to find ideal exponents that maximize the likelihood $\mathcal{L}$ (\textbf{Fig. 1c}) \cite{virkar2014power}. In our results, exponents are estimated as $\alpha=1.69$ ($\upsilon=7.13\%$), $\beta=1.32$ ($\upsilon=-4.53\%$), $\beta_{-}=1.32$ ($\upsilon=-5.06\%$), and $\beta_{+}=1.34$ ($\upsilon=-14.88\%$), where $\upsilon<10\%$ is a reasonable standard for the goodness of ideal estimations. Our estimated power-law models are shown in \textbf{Fig. 1d}. Please see \ref{AP4} for details of power-law exponent estimation.

\begin{figure*}[b]%
\centering
\includegraphics[width=1\textwidth]{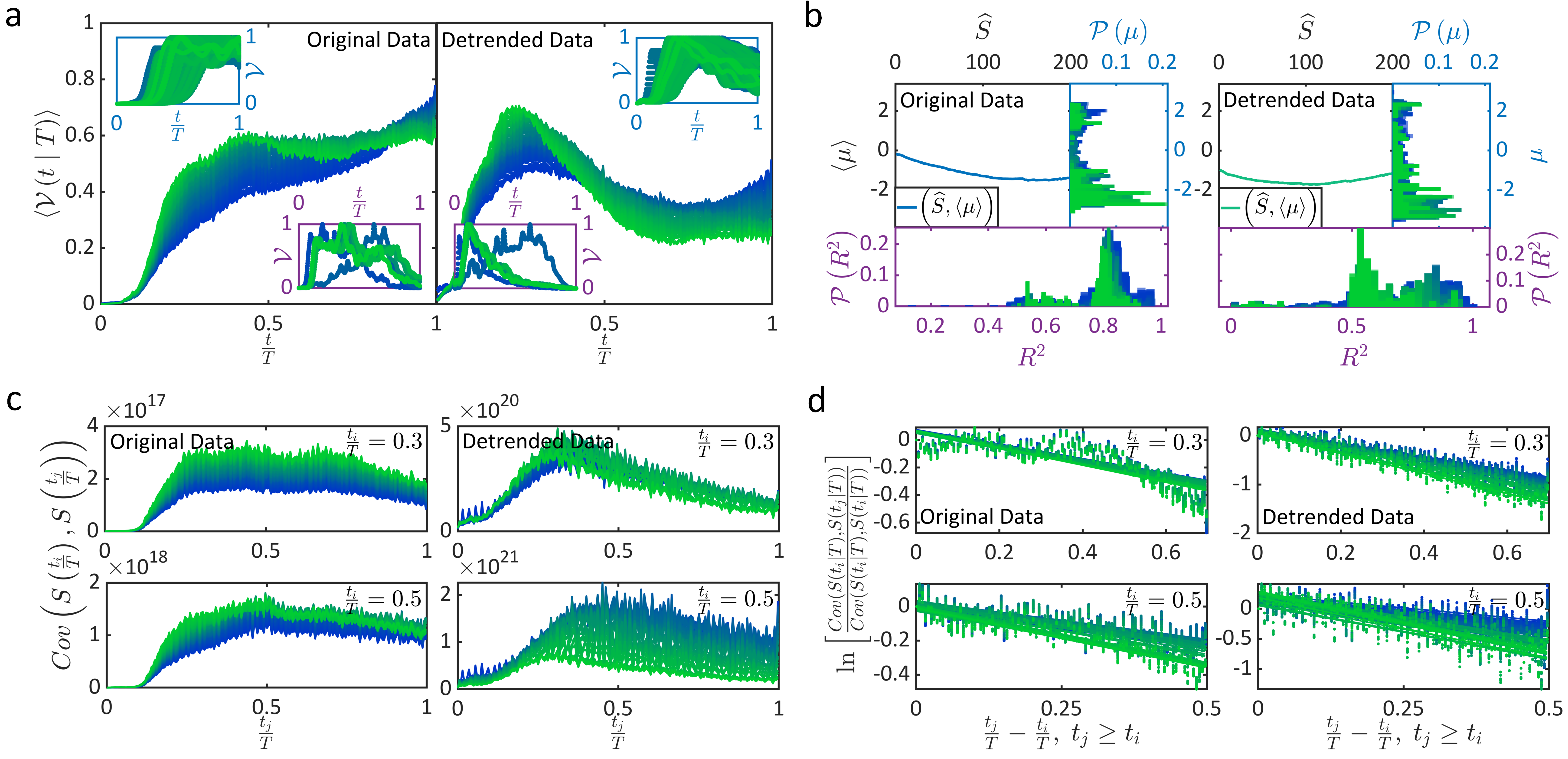}
\caption{Criticality of SARS-CoV-2 evolution. \textbf{a} The collapse shape $\langle\mathcal{V}\left(t\mid T\right)\rangle$ vs. $\frac{t}{T}$ of original and detrended data are shown (main plots), accompanied by several examples of $\mathcal{V}\left(t\mid T\right)$ vs. $\frac{t}{T}$ (inserted small plots, blue boxes correspond to avalanches with $T\in\left[50,200\right]$ and $S\geq 50S^{\prime}$ while purple boxes correspond to avalanches with $T\geq 400$ and $S\geq 50S^{\prime}$). \textbf{b} The average quadratic coefficient $\langle \mu\rangle$ shown as a function of $\widehat{S}$ (black boxes), the probability distribution of $\mu$ (blue boxes), and the $R^2$ statistics of curve fitting (purple boxes). \textbf{c-d} The auto-correlation functions of original and detrended data and their exponential decays.}\label{fig2}
\end{figure*}
 
 Meanwhile, we implement least square fitting on $\langle S\rangle\left(T\right)$ vs. $T$ to derive $\gamma_{+}=1.8431$ ($R^2=0.9112$), $\gamma=1.9363$ ($R^2=0.8871$), and $\gamma_{-}=1.9944$ ($R^2=0.8652$) in \textbf{Fig. 1b}. Together with our estimated avalanche exponents $\alpha$ and $\beta$, these results coincide with the scaling relation in Eq. (\ref{EQ4}), suggesting potential criticality.
 
 To analyze criticality more precisely, we calculate collapse shape following Refs. \cite{baldassarri2003average,pausch2020time}. Please see details of collapse analysis in \ref{AP3}. To ensure the robustness of our results, we further verify if the estimators of distribution cutoffs for $T$ and $S$ affects collapse shape. In \textbf{Fig. 2a}, we show $\langle\mathcal{V}\left(t\mid T\right)\rangle$ vs. $\frac{t}{T}$ of avalanches with $T\geq T^{\prime}$ and $S\geq \widehat{S}$, where $\widehat{S}\in\{S^{\prime},\ldots,200S^{\prime}\}$ (the main panel of \textbf{Fig. 2a}). By increasing $\widehat{S}$, we can simultaneously increase the distribution cutoff of $T$. Our results suggest that the trend of $\langle\mathcal{V}\left(t\mid T\right)\rangle$ is principally independent of $\widehat{S}$. Meanwhile, we notice that $\langle\mathcal{V}\left(t\mid T\right)\rangle$ exhibits a mixture of global increasing trend and parabolic trend. The increasing trends may reflect the intrinsic property of the raw data. Therefore, we recalculate $S$ by multiplying the midpoint of confidence interval \cite{database} with the average total number of worldwide patients in the WHO database \cite{whodatabase} for detrending. In \textbf{Fig. 2a}, $\langle\mathcal{V}\left(t\mid T\right)\rangle$ plausibly exhibits a parabolic trend in the detrended data. Meanwhile, we show several instances of $\mathcal{V}\left(t\mid T\right)$ vs. $\frac{t}{T}$ in \textbf{Fig. 2a}. In \textbf{Fig. 2b}, the parabolic trend of $\langle\mathcal{V}\left(t\mid T\right)\rangle$ is verified quantitatively by quadratic polynomial fitting. We show the average quadratic coefficient $\langle\mu\rangle$ (averaged across avalanches under each condition of $\widehat{S}$) as a function of $\widehat{S}$. Consistent with \textbf{Fig. 2a}, $\langle\mu\rangle<0$ holds for every $\widehat{S}$ in both original and detrended data, suggesting plausible parabolic trends of $\langle\mathcal{V}\left(t\mid T\right)\rangle$. The probability distributions of $\langle\mu\rangle$ and the goodness of fitting, $R^{2}$, are also shown.
 
 In \textbf{Fig. 2c}, we show the auto-correlation (covariance) $\operatorname{Cov}\left(S\left(t_{i}\mid T\right),S\left(t_{j}\mid T\right)\right)$ under each condition of $\widehat{S}$, where we select $t_{i}\in\{0.3T,0.5T\}$ as instances. Please see \ref{AP3} for details. In \textbf{Fig. 2d}, the corresponding decays of auto-correlation are fitted to derive the decay rate $\xi$. In the raw data, we obtain $\langle\xi\rangle=-0.5407$ (averaged $R^2=0.6301$) for $t_{i}=0.3T$ and $\langle\xi\rangle=-0.5159$ (averaged $R^2=0.6324$) for $t_{i}=0.5T$. In the detrended data, we find that $\langle\xi\rangle=-1.6679$ (averaged $R^2=0.6301$) for $t_{i}=0.3T$ and $\langle\xi\rangle=-1.4251$ (averaged $R^2=0.6332$) for $t_{i}=0.5T$. These results suggest exponential decays of auto-correlation with reasonable errors, offering a practical verification of potential criticality.
 
 \begin{figure}[b]%
\centering
\includegraphics[width=0.8\textwidth]{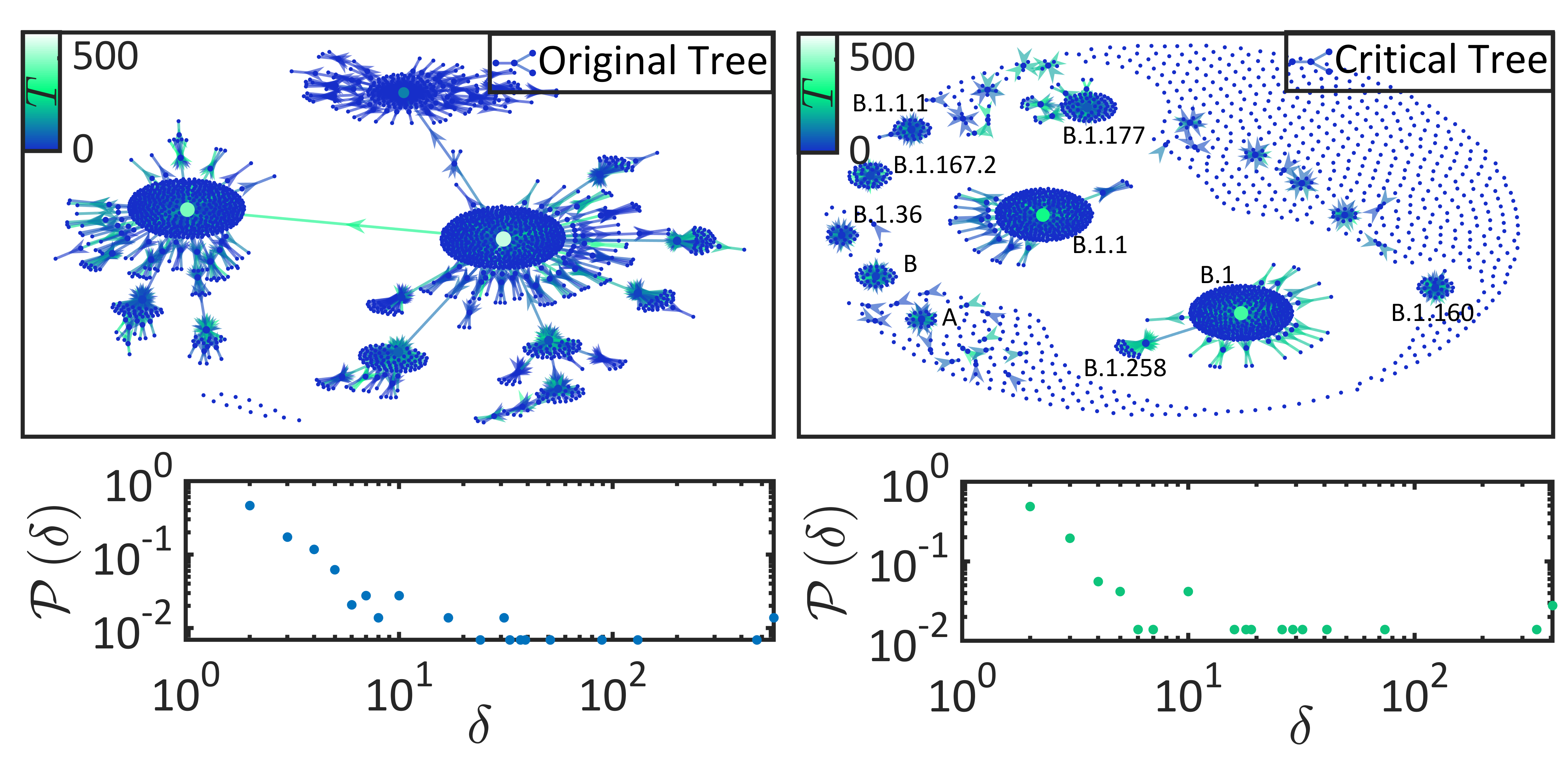}
\caption{Original and critical phylogenetic trees. Edges are colored according to $T$ of corresponding avalanches.}\label{fig3}
\end{figure}
 
 Interestingly, our results coincide with threatening lineages of SARS-CoV-2 if we filter directed edges in its original phylogenetic tree (edges correspond to avalanches and indicate evolution paths of lineages) as following: an edge is removed unless its avalanche satisfies $S\geq S^{\prime}$ and $T\geq T^{\prime}$ (\textbf{Fig. 3}). The filtered result, referred to as critical phylogenetic tree, is a sub-graph of the original one that includes only critical avalanches. Threatening lineages, such as B.1.167.2 (the Delta virus), B.1.1 (the parent of Alpha, Beta, and Omicron viruses), and B.1 (the parent of Epsilon, Eta, Iota, and Mu viruses), can be found by searching non-isolated lineages with high out-degree $\delta$ (number of offspring lineages) in the critical tree. Note that the out-degree in the original tree does not completely determine the possibility of being isolated in the critical tree. Numerous lineages with high out-degrees in the original tree still become isolated (e.g., $20.83\%$, $8.33\%$, and $9.09\%$ of lineages with $\delta\geq5$, $\delta\geq15$, and $\delta\geq20$ in the original tree become isolated after filtering). 
 
 \section{Conclusion}
Following a clue that host cell (human host) resources shape RNA virus mutation and replication, we have naturally derived a theory to characterize RNA virus evolution as a system with absorbing states and avalanches. Our framework maps fundamental biological factors (e.g., phylogenetic tree and infection) to general theories of directed percolation and self-organized criticality, identifying potential physics laws of RNA virus evolution directly from accessible empirical data. Compared with the evolution characterized from the perspective of population dynamics \cite{fontana1989physical,tsimring1996rna,sasaki1994evolution,haraguchi1997evolutionary,sasaki2000antigenic,nowak2000virus,kauffman1993origins,korobeinikov2004global,regoes1998virus,adiwijaya2010multi,garcia2018clonal,yan2019phylodynamic,held2019survival,rasmussen2019coupling,marchi2021antigenic,korobeinikov2016multi}, our framework may be more applicable to analyzing the macroscopic characteristics of evolution dynamics underlying RNA virus phylogenetic tree formation \cite{lassig2014adaptive,neher2014predicting,desai2007beneficial,rouzine2003solitary}.

Applying this theory, we discover that SARS-CoV-2 evolution may proceed at self-organized criticality. Threatening lineages (e.g., the Delta virus) can be coincidentally found on the critical phylogenetic tree. These findings suggest the potential of our theory in analyzing RNA virus evolution and related phenomena. Although the present framework has been validated on SARS-CoV-2, its generalizability is suggested to be further verified on other RNA viruses. Moreover, future studies can also investigate the relation between the self-organized critical evolution process and the threatening degree of RNA virus. Hopefully, this direction may help identify potential virus lineages to be concerned from a physically-explainable perspective.  

\section*{Acknowledgements}
Correspondence should be addressed to Y.T. and P.S. Authors Z.R.T. and H.D.H. contribute equally to this work. This project is supported by the Artificial and General Intelligence Research Program of Guo Qiang Research Institute at Tsinghua University (2020GQG1017) as well as the Tsinghua University Initiative Scientific Research Program. 

 \begin{appendix}
 \section{Host cell resources shape RNA virus evolution}\label{AP1}
With reasonable errors, the mutation rate $\sigma$ of a single RNA virus is typically 
\begin{align}
    \sigma\in\left[10^{-5}\chi\phi\frac{1}{\tau},10^{-4}\chi\phi\frac{1}{\tau}\right],\label{MEQ1}
\end{align}
where $\chi\in\mathbb{N}^{+}$ measures the number of nucleotides, parameter $\tau$ denotes entrance and eclipse periods, and $\phi\simeq 10^{3}$ denotes burst size (the number of viruses produced by a host cell) \cite{domingo1996basic,tsimring1996rna,bar2020science}. These parameters are principally constant for a given RNA virus. Note that the original RNA virus vanishes after replication \cite{v2021coronavirus}. 

If host cell resources are sufficient, RNA viruses may have an opportunity to replicate independently and freely. When these happen, a supremum of the cumulative mutation rate of an RNA virus population of initial size $\omega\in\mathbb{N}^{+}$ during a duration of $\left[0,\kappa\right]$ (assume $\tau$ is divisible  by $\kappa$) can be reached
\begin{align}
    \sup_{\omega}\zeta\left(\omega\right)=&\omega\sigma\int_{0}^{\kappa}\Big[\phi^{\lfloor t/\tau\rfloor}-\phi^{\lfloor t/\tau\rfloor-1}\Big] dt,\label{MEQ2}\\
    =&\omega\sigma\tau\Big(\phi^{\lfloor \kappa/\tau\rfloor-1}-\frac{1}{\phi}\Big),\label{MEQ3}
\end{align}
where notion $\lfloor\cdot\rfloor$ denotes the floor function. 

However, these ideal conditions are not realistic since host cell resources are always limited. Therefore, biologically sound models, including the discussed previous works \cite{nowak2000virus,korobeinikov2004global,adiwijaya2010multi,yan2019phylodynamic,held2019survival,marchi2021antigenic,korobeinikov2016multi}, of RNA virus population dynamics bound the proliferation of RNA virus by time-dependent host cell resource $\lambda\left(t\right)$ (e.g., the number of susceptible cells)
\begin{align}
    &\sup_\omega\widehat{\zeta}\left(\omega\right)= \omega\sigma\int_{0}^{\kappa}\min\Big[\lambda\left(t\right),\phi^{\lfloor \frac{t}{\tau}\rfloor}-\phi^{\lfloor \frac{t}{\tau}\rfloor-1}\Big] dt.\label{MEQ4}
\end{align}
Although Eq. (\ref{MEQ4}) can not be solved until $\lambda\left(t\right)$ is given, we can know
\begin{align}
    \sup_{\lambda}\sup_\omega\widehat{\zeta}\left(\omega\right)=\sup_\omega\zeta\left(\omega\right).\label{MEQ5}
\end{align}
The difference $\sup_\omega\zeta\left(\omega\right)-\sup_\omega\widehat{\zeta}\left(\omega\right)$ will be minimized when $\lambda\left(t\right) \geq \phi^{\lfloor t/\tau\rfloor}-\phi^{\lfloor t/\tau\rfloor-1}$.

\section{Avalanches of RNA virus evolution}\label{AP2}
Consider a time-continuous infection process, where a random patient at moment $t\in\left[t^{\prime},t^{\prime\prime}\right]$ implies three possibilities: becoming absorbed with probability $\rho$, creating another patient with probability $\theta$, or remaining effect-free with probability $1-\left(\rho+\theta\right)$. In critical states, we have $\rho=\theta$ \cite{garcia1993branching}. We define $\mathcal{A}_{n}\left(t\right)$ as the probability for $n$ patients to exist at $t^{*}+t$ given that $1$ patient exists at $t^{*}$. Assuming the independence of patient emergence, we have 
\begin{align}
    \mathcal{A}_{n}\left(t\right)=\sum_{\substack{n_{1}+\ldots+n_{k}=n}}\mathcal{A}_{n_{1}}\left(t\right)\ldots\mathcal{A}_{n_{k}}\left(t\right).\label{MEQ6}
\end{align}
Assume that $\mathcal{A}_n(t),n\in \mathbb{N}^{+}$ admits a Maclaurin expansion $\mathcal{A}_{n}\left(t\right)=a_{n}t+o\left(t^2\right)$ (when $n\neq 1$) or $\mathcal{A}_{n}\left(t\right)=a_{n}t+1+o\left(t^2\right)$ (when $n=1$) where $a_{n}=\mathrm{d}\mathcal{A}_{n}\left(0\right)/\mathrm{d}t$, we can readily derive $a_{0}=a_{2}=\rho$ and $a_{1}=-2\rho$ \cite{garcia1993branching}. Meanwhile, we can know
\begin{align}
\mathcal{A}_{n}\left(t+\mathrm{d} t\right)-\mathcal{A}_{n}\left(t\right)=\sum_{k=0}^{\infty}a_{k}\mathcal{A}_{n-k}\left(t\right)\mathrm{d}t.\label{MEQ7}
\end{align}
Eqs (\ref{MEQ6}-\ref{MEQ7}) readily lead to
\begin{align}
\frac{\partial}{\partial t}\mathcal{G}\left(x,t\right)=&\sum_{k=0}^{\infty}a_{k}\sum_{n=0}^{\infty}\left(\sum_{\substack{n_{1}+\ldots+n_{k}=n-k}}\prod_{i=1}^{k}\mathcal{A}_{n_{i}}\left(t\right)\right)x^{n},\label{MEQ8}\\
=&\sum_{k=0}^{\infty}a_{k}\mathcal{G}\left(x,t\right)^{k},\label{MEQ9}
\end{align}
where $\mathcal{G}\left(x,t\right)=\sum_{n=0}^{\infty}\mathcal{A}_{n}\left(t\right)x^{n},\;x\in\left[0,1\right]$ denotes the generating function. Applying a trick introduced in Ref. \cite{garcia1993branching}, we define 
\begin{align}
    \mathcal{H}\left(x\right)=\frac{\partial}{\partial t}\mathcal{G}\left(x,0\right),\label{MEQ10}
\end{align}
which naturally leads to 
\begin{align}
    \frac{\partial}{\partial t}\mathcal{G}\left(x,t\right)=\mathcal{H}\left(\mathcal{G}\left(x,t\right)\right),\label{MEQ11}
\end{align}
the backward Chapman-Kolmogorov equation. Meanwhile, $\mathcal{H}\left(x\right)=\rho\left(1-x\right)^{2}$ can be derived based on $a_{0}$, $a_{1}$, and $a_{2}$ \cite{garcia1993branching}. Taken together, we have
\begin{align}
    \frac{\partial}{\partial t}\mathcal{G}\left(x,t\right)=\rho\left(1-\mathcal{G}\left(x,t\right)\right)^{2}.\label{MEQ12}
\end{align}
Note that the initial condition is $\mathcal{G}\left(x,0\right)=x$ since one patient emerges at $t^{*}$. Solving Eq. (\ref{MEQ12}), we derive an analytic expression
\begin{align}
    \mathcal{G}\left(x,t\right)=\frac{\rho\left(1-x\right)t}{\rho\left(1-x\right)t+1}.\label{MEQ13}
\end{align}
Therefore, we have $\mathcal{A}_{0}\left(t\right)=\mathcal{G}\left(0,t\right)=\frac{\rho t}{\rho t+1}$, supporting a calculation of lifetime distribution $\mathcal{P}_T\left(t\right)$
\begin{align}
    \lim_{t\rightarrow\infty}\mathcal{P}_T\left(t\right)= \lim_{t\rightarrow\infty}\frac{\mathrm{d}}{\mathrm{d}t}\mathcal{G}\left(0,t\right)\sim t^{-2}.\label{MEQ14}
\end{align}
Following Refs. \cite{garcia1993branching,harris1963theory,otter1949multiplicative}, one can similarily calculate 
\begin{align}
    \lim_{s\rightarrow\infty}\mathcal{P}_S\left(s\right)\sim s^{-\frac{3}{2}}.\label{MEQ15}
\end{align}
These exponents are consistent as the predictions of other mean field theories \cite{zapperi1995self,garcia1993branching}.

\section{Verify the criticality of RNA virus evolution}\label{AP3}
Compared with the scaling relation \cite{sethna2001crackling,baldassarri2003average,hinrichsen2000non,lubeck2004universal}, a more precise verification of criticality can be implemented by the collapse shape \cite{baldassarri2003average,pausch2020time}
\begin{align}
    \langle\mathcal{V}\left(t\mid T\right)\rangle=\mathcal{F}\left(\frac{t}{T}\right),\label{MEQ17}
\end{align}
in which the expectation is averaged across different $T$. Notion $\mathcal{F}\left(\cdot\right)$ denotes a universal scaling function. Notion $\mathcal{V}\left(t\mid T\right)$ is defined as the average collapse shape of all avalanches with the same life time $T$ 
\begin{align}
    \mathcal{V}\left(t\mid T\right)= T^{1-\gamma}\langle S\left(t\mid T\right)\rangle.\label{MEQ18}
\end{align}
where $S\left(t\mid T\right)$ denotes the time-dependent avalanche size during these avalanches. At criticality, all data of $\mathcal{V}\left(t\mid T\right)$ should collapse onto $\mathcal{F}\left(\cdot\right)$, a parabolic function, with reasonable errors \cite{baldassarri2003average,pausch2020time}. In some cases, function $\mathcal{F}\left(\cdot\right)$ may contain both the parabolic component and a slight global trend (e.g., increasing or decreasing), making the shape not perfectly parabolic. De-trending can be used to deal with this issue during data pre-processing. 

Moreover, a more practical verification concerns the slow and exponential decay of auto-correlation \cite{pausch2020time}
\begin{align}
    \ln\left[\frac{\operatorname{Cov}\left(S\left(t_{i}\mid T\right),S\left(t_{j}\mid T\right)\right)}{\operatorname{Cov}\left(S\left(t_{i}\mid T\right),S\left(t_{i}\mid T\right)\right)}\right]=-\xi\left(\frac{t_{j}-t_{i}}{T}\right)+c,\label{MEQ19}
\end{align}
where $t_{i}\in\left[t^{\prime},t^{\prime\prime}\right]$ and $t_{j}\in\left[t_{i},t^{\prime\prime}\right]$.

\section{Power-law analysis}\label{AP4}

The probability distributions of life time $T$ and avalanche size $S$ (for midpoint and bounds) are derived after a fine-grained data binning process with $1000$ bins is applied, which is useful in de-noising while controlling information loss. 

We discover that $\mathcal{P}_{T}\left(\cdot\right)$ does not follow power-law properties for all values of $T$ (see \textbf{Fig. 1b}). Therefore, distribution cutoff estimation is necessary for $T$ and $S$. We apply the method introduced in Ref. \cite{virkar2014power} to estimate $T^{\prime}$, $S^{\prime}$, $S_{-}^{\prime}$ (lower bound), and $S_{+}^{\prime}$ (upper bound), which calculates the Kolmogorov-Smirnov (KS) statistic $\eta$ between the observed cumulative probability distribution above cutoff and the cumulative probability distribution of a standard power-law variable. An ideal cutoff is expected to minimize $\eta$ \cite{virkar2014power}. To control the reduction of sample size, we relatively relax the restriction and estimate cutoffs when $\langle\eta\rangle-0.5\operatorname{std}\left(\eta\right)$ is reached at the first time, where $\operatorname{std}\left(\cdot\right)$ denotes the standard deviation. Distribution cutoffs are estimated as $T^{\prime}=29$, $S^{\prime}=13320$, $S_{-}^{\prime}=12760$, and $S_{+}^{\prime}=13980$. The distribution tails above cutoffs cover $93.879\%$,  $76.38\%$, $70.526\%$, and $91.683\%$ of original samples, supporting robust estimations of power-law exponents. 

A maximum likelihood estimation of power-law exponent \cite{virkar2014power} is implemented on the samples above cutoffs. An ideal exponent can maximize the likelihood $\mathcal{L}$ \cite{virkar2014power}. After estimating power-law exponents, we carry out a KS-statistic-based analysis to derive the goodness of estimation. To calculate the average KS statistic $\eta^{*}$ between sample distributions and estimated models, we first generate $1000$ sample distributions (each sample distribution contains $n\in\left[500,5000\right]$ samples) of estimated power-law models. Then, we define $\upsilon=\frac{\widehat{\eta}-\eta^{*}}{\eta^{*}}$ to reflect the goodness of estimation, where $\widehat{\eta}$ is the KS statistic between the cumulative probability distributions of estimated power-law models and empirical distributions above cutoffs. We suggest that $\upsilon<10\%$ can be a reasonable standard for ideal estimations.
 \end{appendix}
 
\section*{References}
\bibliographystyle{iopart-num}
\bibliography{ref}
\end{document}